# Strain Modulation of Graphene by Nanoscale Substrate Curvatures: A Molecular View

Yingjie Zhang,[†,‡,§] Mohammad Heiranian,[†,‖] Blanka Janicek,[⊥] Zoe Budrikis,[#] Stefano Zapperi,[#,¶,△,▼] Pinshane Y. Huang,[⊥] Harley T. Johnson,[‖] Narayana R. Aluru,[†,‖] Joseph W. Lyding,[†,‡] and Nadya Mason*,[§]

[†]Beckman Institute for Advanced Science and Technology, [‡]Department of Electrical and Computer Engineering, [§]Department of Physics and Frederick Seitz Materials Research Laboratory, [‖]Department of Mechanical Science and Engineering and [⊥]Department of Materials Science and Engineering, University of Illinois, Urbana, Illinois 61801, United States

[#]ISI Foundation, Via Chisola 5, 10126 Torino, Italy

[¶]Center for Complexity and Biosystems, Department of Physics, University of Milano, Via Celoria 16, 20133 Milano, Italy

[△]CNR - Consiglio Nazionale delle Ricerche, Istituto di Chimica della Materia Condensata e di Tecnologie per l'Energia, Via R. Cozzi 53, 20125 Milano, Italy

[▼]Department of Applied Physics, Aalto University, P.O. Box 11100, FI-00076 Espoo, Finland

Ⓢ *Supporting Information*

**ABSTRACT:** Spatially nonuniform strain is important for engineering the pseudomagnetic field and band structure of graphene. Despite the wide interest in strain engineering, there is still a lack of control on device-compatible strain patterns due to the limited understanding of the structure−strain relationship. Here, we study the effect of substrate corrugation and curvature on the strain profiles of graphene via combined experimental and theoretical studies of a model system: graphene on closely packed $SiO_2$ nanospheres with different diameters (20−200 nm). Experimentally, via quantitative Raman analysis, we observe partial adhesion and wrinkle features and find that smaller nanospheres induce larger tensile strain in graphene; theoretically, molecular dynamics simulations confirm the same microscopic structure and size dependence of strain and reveal that a larger strain is caused by a stronger, inhomogeneous interaction force between smaller nanospheres and graphene. This molecular-level understanding of the strain mechanism is important for strain engineering of graphene and other two-dimensional materials.

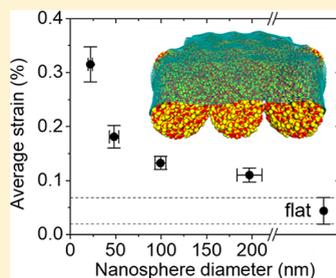

T wo-dimensional materials are promising for next-generation electronics, due to their versatile band structure, high mobility, and superior electric field-effect tunability.[1−4] Atomically thin in nature, these materials also exhibit high mechanical flexibility, enabling mechanical tuning of electronic properties.[5−12] Among the rich library of 2D materials, graphene shows the highest electronic mobility to date, although its semimetallic nature presents a significant obstacle for device applications.[2,4] Mechanical strain of graphene offers a route to overcome this obstacle via band engineering.[13−15] While gap opening can also be achieved via size confinement in the form of nanoribbons or quantum dots, strain engineering enables continuous tuning of graphene's electronic structure and gap without introducing edge defects and provides a platform for novel effects such as pseudomagnetic field generation, zero-field quantum Hall states, and topological valley Hall transport.[13−15] However, band structure engineering in graphene requires spatially nonuniform strain, which is hard to achieve using conventional techniques such as epitaxial lattice mismatch and mechanical stretching.[13−17] Controllable, device-compatible nonuniform strain patterns in graphene can be engineered by depositing graphene on corrugated

substrates.[18−21] Although different strain patterns have been demonstrated using this approach, the microscopic mechanism of the correlation between substrate corrugation features (height profile, characteristic size, curvature, etc.) and the strain profiles of graphene is still elusive. This lack of microscopic understanding is a bottleneck for designing and fabricating strained graphene devices and for observing strain modulated electronic transport in devices.

To examine the strain mechanisms, we systematically vary substrate corrugation features and study the evolution of deformation patterns and the magnitude of strain induced in graphene (Gr). We choose periodic spherical curvature patterns in the form of closely packed nanospheres, which enable controllable diameter/curvature tuning. Previous studies have demonstrated deformation properties of Gr on isolated $SiO_2$ nanoparticles[22] and Gr on multilayer nanoparticle films where the top layer of particles is randomly spaced.[20,21] Detailed discussions of these prior results can be found in Supporting

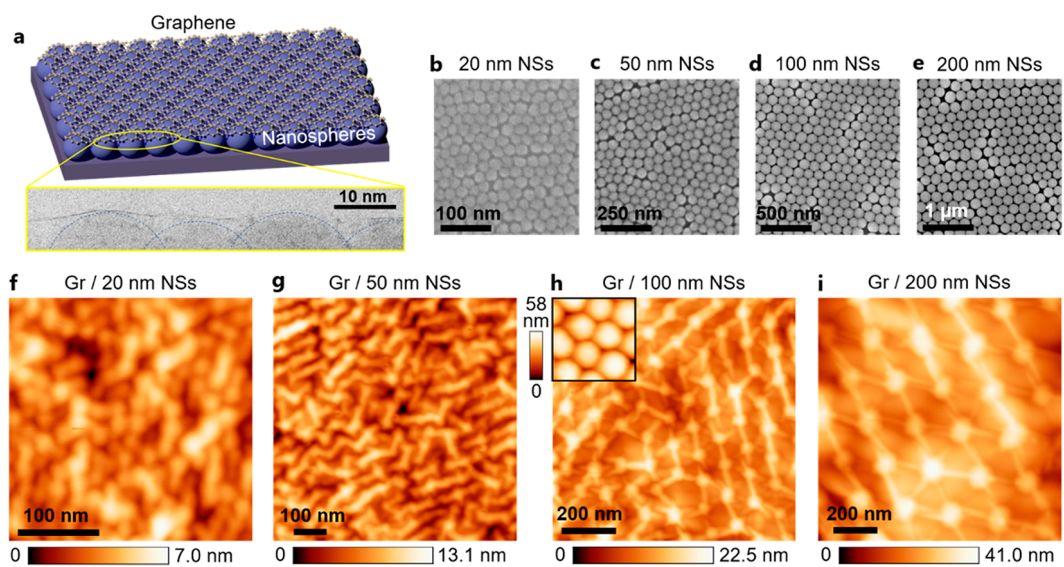

**Figure 1.** Structure of the graphene on nanosphere systems. (a) Top panel: schematic showing graphene deposited on monolayers of hexagonal close-packed $SiO_2$ nanospheres. Bottom panel: cross-sectional bright-field scanning transmission electron microscopy image of graphene on 20 nm NSs. Graphene is seen as a black curved line on top of the NSs. Blue dashes mark the boundary of the NSs. (b–e) Scanning electron microscopy images of the NS monolayers before graphene deposition. (f–i) Corresponding atomic force microscopy (AFM) images of Gr on NSs. NS diameters are labeled in (b–i). The top-left inset in (h) is an AFM image of 100 nm NSs with no graphene on top, showing larger height variations compared to that with the Gr-covered regions. The contrast between the structure of the bare NS assemblies (b–e) and the Gr-NS samples (f–i) highlights the deformation and wrinkle structures of Gr induced by the underlying NSs.

Information, section 6. In comparison, our system consists of monolayers of spherical particles with controlled, minimum spacing due to the close-packing properties and is thus ideal for use as a model system to study the correlation mechanisms between substrate curvature and strain. Combined with molecular dynamics (MD) simulations, we found unequivocally that larger tensile strain is induced in systems with smaller local radius of curvature (with close-packed curvature patterns). We attribute this to spatial force inhomogeneities at the molecular level, which is a critical mechanism underlying strain modulations that has been overlooked in previous theoretical analyses.[22−24]

### ■ EXPERIMENTAL RESULTS

We prepared monolayers of hexagonal close-packed $SiO_2$ nanospheres (NSs) on $SiO_2$/Si substrates (Supporting Information, sections 1 and 2) and deposited single-layer graphene on top (Figure 1a). $SiO_2$ NSs are chosen due to their well-controlled spherical shape, wide range of available sizes (20–400 nm), insulating nature (desired for electronic device fabrication), and clean surfaces (no organic surface ligands). When deposited on substrates, NSs assemble together, forming separated monolayer regions having sizes ranging from a few microns to tens of microns; between neighboring monolayer regions, there are usually very few NSs (Supporting Information, Figure S1). After depositing graphene on top of the NS substrate, we typically observe unstrained, flat graphene regions (on the flat substrate) adjacent to strained Gr areas (on the NS monolayer domains). The coexistence of strained and unstrained regions near each other enables control studies to rule out the effect of sample-to-sample variations on the experimental results.

For experimental studies, we chose four different NS diameters: 20 nm (22.0 ± 2.3 nm), 50 nm (48.1 ± 5.3 nm), 100 nm (99.2 ± 5.8 nm), and 200 nm (197.1 ± 13.5 nm). A cross-sectional bright-field scanning transmission electron microscopy (STEM) image of the Gr/20 nm NSs system (Gr-20, similar abbreviations apply for other sizes) is shown in Figure 1a (bottom), revealing deformation and partial adhesion of graphene on the NSs. Scanning electron microscopy (SEM) images of the monolayer NS assemblies (before Gr deposition) are shown in Figure 1b–e, revealing hexagonal close-packing order. Although dislocations and grain boundaries exist, hexagonal ordering typically extends over more than tens or hundreds of NSs (Figure S2). Atomic force microscopy (AFM) images of the Gr on NS systems (Gr-NS) (Figure 1f–i) reveal the height and deformation profile of graphene. In contrast to the bare NS assemblies, where the spatial pattern is independent of NS size, graphene deposited on these NSs show different deformation profiles as the NS diameter changes. While Gr-20 exhibits (partial) conformal coating, systems with larger spheres show wrinkle patterns of graphene connecting neighboring NSs. For Gr-100 and Gr-200, we can most clearly see that graphene adheres only to the apex of the spheres and is free-standing between neighboring NSs. Wrinkle features in Gr/nanoparticle systems have been observed before, as discussed in Supporting Information, section 6. The wrinkle formation in our systems is due to the geometrical frustration of the graphene membrane on the spherical substrate, together with the competition between the graphene−substrate adhesion energy and the internal strain energy of graphene. While the wrinkle effects have been explained by continuum mechanics models,[22−24] these models are not sufficient to explain the size dependence of strain as we will show later. Note that these partial adhesion and wrinkle effects are different from the "snap-through" transition effects observed before,[25−27] which only occur for thick, multilayer graphene that has strong bending stiffness.

Although the absolute values of the Gr adhesion area on top of each NS becomes larger as the NSs become larger, the width of wrinkles is nearly the same, in the range of 15–20 nm for Gr-50, Gr-100, and Gr-200. This is possibly due to a balance

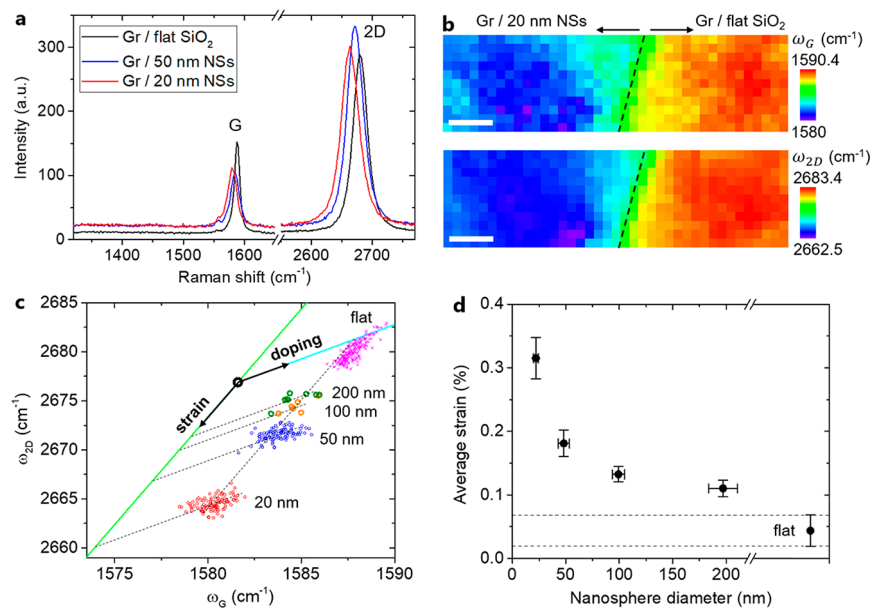

**Figure 2.** Strain quantification via confocal Raman spectroscopy. (a) Raw Raman spectrum of flat Gr, Gr-20, and Gr-50. (b) G and 2D peak position map of an area containing Gr-20 on the left and flat Gr on the right, where the boundary is marked by the dashed lines. Scale bars: 1 $\mu$m. (c) Correlation map of the G and 2D peak positions for different systems as labeled. For Gr-20 and Gr-50, each data point represents a spectrum taken over an area of ~0.5 × 0.5 $\mu$m$^2$; for Gr-100 and Gr-200, the Raman signal is weaker and each data point corresponds to a ~2 × 2 $\mu$m$^2$ region. The black circle at (1581.6, 2676.9) represents a perfect graphene system having zero doping and strain. The green and cyan lines represent the strain and doping directions. Detailed analysis is shown in Supporting Information, section 3.2. (d) Spatially averaged areal strain as a function of NS diameter, extracted from the peak correlation maps. Each of the strain values is averaged over multiple areas in multiple samples to ensure statistical significance. The right-most data point represents flat Gr, where the small tensile strain is induced by the angstrom-level corrugation of the "flat" SiO$_2$ substrate.[29−31]

between the tension force in Gr and the friction force between Gr and NS.[28] Phenomenologically, this fixed width of wrinkles is responsible for the evolution of Gr deformation patterns as the NS size increases: for Gr-20, the NSs are too small to generate wrinkles, and therefore, the Gr experiences smooth deformations (Figure 1f); in Gr-50, the wrinkle width is similar to the diameter of the adhered Gr area on each NS, facilitating the propagation of wrinkles and the formation of stripy domains, where each NS is typically connected to four of its six neighboring spheres via Gr wrinkles (Figure 1g); for larger NSs, the Gr adhesion diameter is larger than the wrinkle width, giving rise to more wrinkles (each NS is typically surrounded by three to six wrinkles connecting to adjacent NSs) (Figure 1h,i).

To quantify the strain effects, we performed confocal Raman spectroscopy on the Gr-NS systems and control samples of Gr on flat SiO$_2$. Typical raw spectra of flat Gr, Gr-20, and Gr-50 are shown in Figure 2a. We note that the laser spot in the Raman measurements has a size of ~0.5 × 0.5 $\mu$m$^2$, covering multiple NSs. Therefore, each measured spectrum contains information on the spatially averaged doping and strain over many NSs. While these average values do not provide the full information on the microscopic strain distribution, it is a reasonable estimate of the overall magnitude of strain. We can see two prominent peaks characteristic of graphene: G mode (1580−1590 cm$^{-1}$) and 2D mode (2660−2680 cm$^{-1}$). The absence of a D peak (~1350 cm$^{-1}$) in all the spectra reveals that the samples are defect-free (within the Raman sensitivity). Compared to flat Gr, Gr-20 and Gr-50 show blue shifts in the G and 2D peaks, with more significant shifts occurring in Gr-20. These blue shifts are directly visualized in Figure 2b, which are peak position maps of a region containing both Gr on nanosphere domains and Gr on flat SiO$_2$ domains (see also Figure S5). At each optical pixel, the measured spectrum is fitted with Lorentzian line shapes for both G and 2D modes, and the fitted peak values ($\omega_G$ and $\omega_{2D}$) are plotted in the map. The large spectral shift and sharp transition across the domain boundary (occurring within ~0.5 $\mu$m, the optical resolution limit) both confirm the strong modulation of the underlying NSs on the Raman scattering of graphene.

It is known that Raman peak shifts in graphene are sensitive to both doping and strain modulations, and these two effects can be deconvoluted by correlation analysis of G and 2D modes.[29] We adopt this analysis method to calculate the spatially averaged areal strain (hydrostatic strain) in graphene. At each optical pixel or NS monolayer domain, the measured peak positions of G mode and 2D mode are plotted as one point in the correlation map shown in Figure 2c; flat Gr, Gr-20, Gr-50, Gr-100, and Gr-200 systems are all included in this correlation plot. Also plotted is a data point (black circle) corresponding to the intrinsic frequencies for doping and strain-free graphene ($\omega_G^0$, $\omega_{2D}^0$) = (1581.6, 2676.9), the line for uniaxially strained graphene (green, slope = 2.2), and the line for charge-doped graphene (cyan, slope = 0.7). Each ($\omega_G$, $\omega_{2D}$) point forms a vector with respect to the origin ($\omega_G^0$, $\omega_{2D}^0$), which can be decomposed into the strain and doping axes. The length of the projected vector along the strain and doping axes is proportional to the spatially averaged areal strain and average hole doping concentration, respectively (Supporting Information, section 3.2). The strain is tensile if the vector projected on the strain axis points down and compressive when the vector points up. The doping concentration of all the systems exhibits spatial fluctuations, but the spatially averaged doping values are of a similar order of magnitude, ~10$^{12}$ cm$^{-2}$ (Figure S3b). In contrast, the average strain of different systems shows a clear increasing trend as the NSs become smaller (Figure 2d). All the systems exhibit an average tensile strain, with a maximum value of 0.32 ± 0.03% for Gr-20. The control sample, flat Gr, shows a small tensile strain of 0.044 ± 0.024% due to the random angstrom-level height variation of the "flat" $SiO_2$ surface, consistent with previous reports.[29−31] All the Gr-NS systems have larger average tensile strain than the flat Gr, revealing that the graphene lattice is stretched when deposited on the NS assemblies. Furthermore, peak width analysis indicates that large nanoscale strain variations are present in Gr-NS systems, and in contrast, local strain fluctuations in flat Gr are much smaller (Supporting Information, section 3.3).

Note that a previous study showed similar trends of Raman shifts for graphene on disordered films of $SiO_2$ nanoparticles.[20] This indicates that, while controllably ordered structures are desired for mechanistic studies, the trend of strain enhancement by smaller substrate radius of curvature is likely a general effect that does not require spatial ordering or periodicity.

**Proposed Curvature Dependence Mechanism.** Previously, analytical continuum mechanics models were developed to explain the strain mechanisms of graphene on a spherical substrate, based on the balance between strain and adhesion energies.[23,24] These models assume uniform Gr-to-NS distance and interaction energy (in the adhered region) and predict that the strain profile and magnitude are independent of the size of the sphere. We propose that our observed size/curvature dependence of strain is due to the molecular-level inhomogeneities of the distance and interaction force between Gr and NSs. On the molecular level, NSs induce strain in graphene due to the van der Waals interaction between them;[32] these interactions are essentially determined by the Lennard-Jones (L-J) potential $V_{LJ} = \frac{A}{r^{12}} - \frac{B}{r^6}$ between the atoms in the Gr and atoms in the NSs. L-J interactions typically extend over a small distance $d_c$ on the order of 1 nm. For a nanosphere with a radius $R$, only a small spherical region of Gr (with radius $r$) feels the interaction with the NS (Figure 3). This interaction range is determined by the limited interaction distance of the L-J potential, $d_c$. We thus have the geometric relation: $r^2 + R^2 = (R + d_c)^2$, from which we obtain $r = \sqrt{2Rd_c + d_c{}^2}$. If $2R \gg d_c$ (i.e., NS radius larger than ∼5 nm), we obtain $r \cong \sqrt{2Rd_c}$. Therefore, the ratio of the graphene area experiencing interactions with the underlying spheres, to the total area of graphene, is roughly $\pi r^2 / (\pi R^2) = 2d_c/R$. Because $d_c$ is a constant distance independent of the sphere radius $R$, we conclude that smaller NSs interact with a larger area fraction of the Gr deposited on top. A more rigorous calculation involving direct integration of the L-J potential over the whole NS, as was done previously,[33] gives the same scaling behavior and confirms that the interaction force per unit area is higher for smaller NS diameter. While our simplified analysis considers the initial state of flat Gr, we expect the same size dependence to persist as the Gr bends and adheres more to the NSs, resulting in higher strain for Gr on top of smaller NSs in the equilibrium state. This will be confirmed by MD simulations shown below. Note that the Gr bending magnitude is much smaller than the diameter of the spheres, as shown in the cross-sectional image in Figure 1a; thus the case of bent Gr is not expected to have large deviations from the proposed approximate calculations for flat Gr. Note that the proposed strain scaling behaviors are applicable to all the experimentally

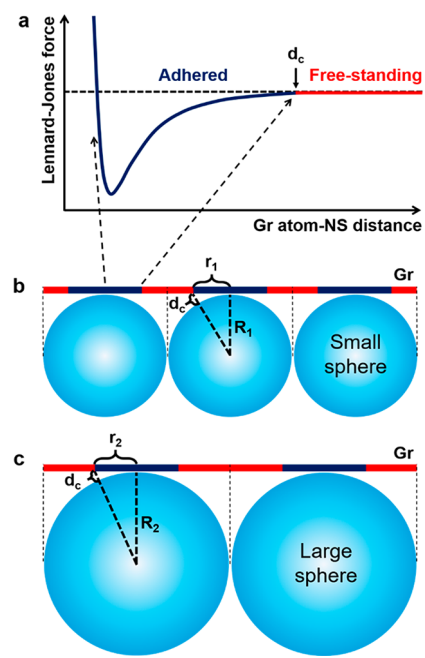

**Figure 3.** Schematic of the microscopic force modulation and strain mechanism for a graphene disk on top of close-packed nanospheres. (a) Schematic of Lennard-Jones (L-J) force versus distance between a graphene (carbon) atom and the underlying sphere. (b,c) Variation of the Gr to NS distance for small and large NSs, respectively. $r_1$ and $r_2$ are the radii of the Gr regions that experience L-J interactions with each underlying NS. $R_1$ and $R_2$ are the radii of the NSs. $d_c$ (∼1 nm, same in (a–c)) is the maximum distance beyond which the L-J force between a graphene atom and the NS is negligible. The regions of Gr interacting and not interacting with NSs are colored dark blue and red, respectively, both in (b) and (c). The same color code is used in the force curve of (a). Following the derivations in the text, we have $\frac{\pi r_1^2}{\pi R_1^2} > \frac{\pi r_2^2}{\pi R_2^2}$; that is, smaller NSs induce larger fractions of Gr-NS interaction area. This model is confirmed by MD simulations for both flat and bent Gr on NSs, as shown in Figure 5.

relevant sphere sizes, ranging from a few nanometers to a few microns.

**Molecular Dynamics Simulations.** To fully examine the molecular-level strain mechanism, we performed MD simulations to study the dynamic process of Gr bending and adhesion to the NSs. Our experimental Gr-NS systems, typically consisting of thousands of NSs where each NS is composed of millions to billions of atoms, is challenging for typical MD simulations. Therefore, we chose four smaller, simpler systems for simulations, where the NS diameters are 5, 10, 20, and 30 nm. Each of the simulated systems consists of a graphene sheet on seven identical closely packed amorphous $SiO_2$ NSs. Figure 4a shows the configuration of the initial state of the 20 nm system where graphene lies flat on top of the spheres (left), and the equilibrium state after the structure of graphene is relaxed (right). Due to the vdW attraction between Gr and the NSs, graphene bends and partially adheres to the NSs during the relaxation process. Figure 4b plots the calculated average areal strain as a function of NS diameter, over the central sphere in each system (see the dashed hexagon in Figure 4f). This central hexagon, a unit cell of the hexagonal close-packed NS array, is chosen to avoid the effect of boundaries on the strain calculations (Supporting Information, Figure S7). Areal strain (hydrostatic strain) is calculated as the change of the local graphene area in the equilibrium state compared to that of the relaxed initial state (Supporting Information, section 4). We find the strain to be tensile, and it becomes larger for smaller NSs; these are both consistent with the experimental results shown in Figure 2d. Moreover, the calculated strain value for Gr-20 is ∼0.36%, close to the experimentally extracted value (0.32%).

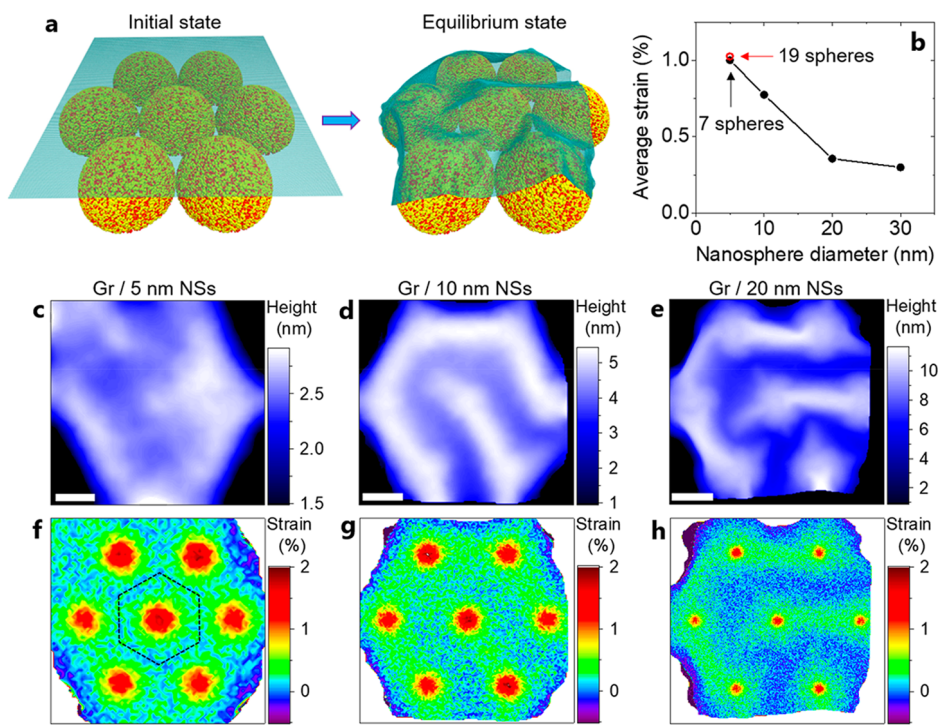

**Figure 4.** Molecular dynamics (MD) simulations of the strain and height profile. (a) 3D configuration of graphene on seven 20 nm NSs in the initial state (flat Gr) and equilibrium state (bent Gr). (b) Average areal strain of Gr on top of the central sphere (region marked as dashed hexagon in (f)), as a function of NS diameter. Black dots represent data obtained from Gr on an assembly of seven NSs, whereas the red circle corresponds to Gr on an assembly of 19 NSs (Supporting Information, Figure S7). The match of the two data points at 5 nm diameter confirms that boundary effects have negligible impact on the obtained strain values. (c−e) Height profiles of Gr on top of 5, 10, and 20 nm NSs, respectively, showing the deformation effects. Scale bars are 2.5, 5, and 10 nm for (c), (d), and (e), respectively. (f−h) Strain profiles of the same systems (and same areas) shown in (c−e), revealing larger overall strain for smaller NSs.

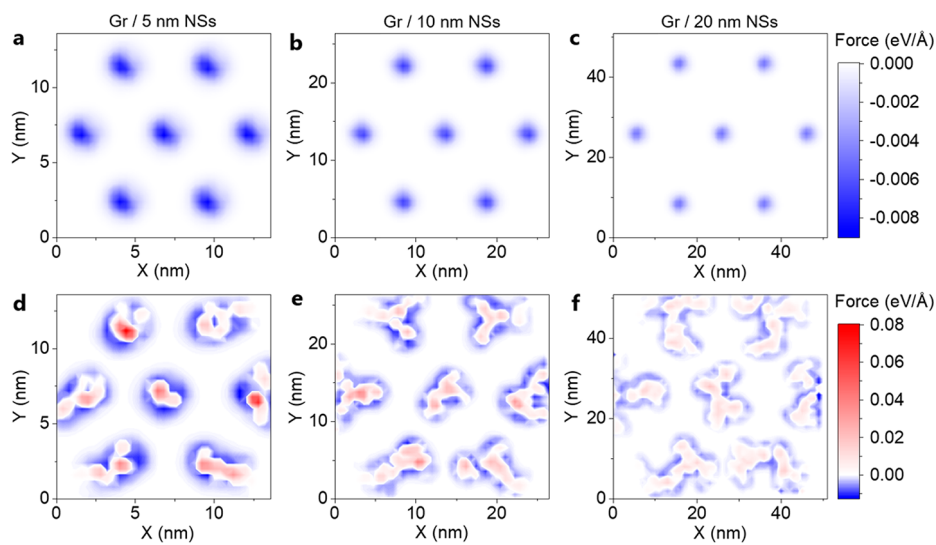

**Figure 5.** MD simulations of the force distributions. (a−c) Force distribution in the initial state (flat Gr) for Gr-5, Gr-10, and Gr-20, respectively, revealing that a larger area fraction of Gr is attracted by smaller NSs. (d−f) Force distribution of the same systems in the equilibrium state (bent Gr), revealing larger forces (both repulsive, in red, and attractive, in blue) between Gr and smaller NSs. Here, the values of the force are obtained as the average out-of-plane force per atom in graphene.

We further calculated the spatial distribution of height and strain of Gr in the equilibrium state, with results shown in Figure 4c−h (and Supporting Information, Figure S6 for the Gr-30 system). The height profile of the smallest simulated system (Gr-5) exhibits smooth deformation with no wrinkles, whereas wrinkles start to form as the size become larger and are pronounced for Gr-20 and Gr-30. This trend agrees with the experimental results in Figure 1f−i, though the threshold NS size for wrinkle formation is slightly larger in experimental systems. The simulated strain distribution maps (Figure 4f−h) reveal maximum tensile strain on the apex of the NSs for all the systems and lowest strain for the free-standing part of graphene between the NSs. In systems having smaller NSs, the high strain areas cover a larger portion of the spheres, and therefore, the average strain is larger. Note that previous experimental results for monolayer $MoS_2$ on ~400 nm nanocones[34] is in qualitative agreement with our simulated strain patterns for Gr-NS (5−30 nm NS diameter): maximum strain at the apex of protrusions and small or zero strain at the valleys. This is a good example revealing that our main conclusions on the strain enhancement mechanisms are likely applicable to a wide variety of 2D materials.

To directly verify the force modulation mechanism we proposed in Figure 3, we calculated the L-J interaction force between the Gr and NSs in the MD simulations. The spatial distribution of the out-of-plane component of the force in the initial and the equilibrium state is shown in Figure 5a−c and d−f, respectively. We find that in the initial state (flat graphene configuration as shown in Figure 4a) the smaller NSs impose attractive force to a larger portion of the graphene sheet, consistent with the cartoon diagram in Figure 3. Due to the larger overall attraction in the smaller NS systems, graphene goes through more significant downward bending during the structural evolution process and therefore becomes more stretched near the apex of the NSs. In the equilibrium (bent and partially adhered) state, graphene experiences a repulsive force at the apex of the spheres in all the systems to balance the local stretching strain, and away from the apex, the force becomes attractive (due to a larger Gr to NS distance) before vanishing at the free-standing part of Gr (Figure 5d−f). The overall repulsive force should be equal to the overall attractive force to ensure force balance, and on top of smaller NSs, the deformed graphene experiences both a larger repulsion at the NS centers and a larger attraction at the peripherals; these larger forces in the equilibrium state are also consistent with a higher overall tensile strain in the system containing smaller NSs.

As the MD simulations are performed mostly on a system containing only seven NSs, it is worth examining the effect of boundaries on the strain modulations. Therefore, we performed a simulation on a system consisting of graphene on 19 close-packed NSs (5 nm diameter). We found that the magnitude and distribution of strain in graphene are similar to that of the seven NS system, in the regions away from the boundaries (Supporting Information, Figure S7). Therefore, we believe the simplified seven NS simulations are good representations of the experimental system containing a large array of NSs.

## ■ DISCUSSIONS

We have studied, via both experiments and simulations, the deformation and strain of graphene on close-packed $SiO_2$ nanosphere arrays and found that strain is enhanced when the NSs are scaled to smaller sizes. The underlying mechanism is the molecular-level distance and force variations between Gr and the NSs, which were not taken into account in previous theoretical strain analyses.[22−24] This is especially important in graphene-corrugated substrate systems where graphene tends to be partially suspended. Because in the equilibrium state strain energy is balanced with the Gr−substrate interaction energy, larger strain requires stronger interaction. If substrate corrugation features are tall and sharp, most of the graphene will be suspended and have zero interaction with the substrate; on the other hand, if the substrate corrugation is smooth and broad over a large area, the Gr−substrate interaction will be uniform and also weak. Therefore, it is most desirable to have spatially connected substrate corrugation features with small radius of curvature (e.g., small NS arrays), so that most of the Gr is attached to, and has strong inhomogeneous interactions with, the substrate. In this way, graphene will have high, nonuniform strain in most areas. This design principle can be applied not only to graphene on $SiO_2$ but also to all other 2D materials on different substrates.

## ■ ASSOCIATED CONTENT

### * Supporting Information

The Supporting Information is available free of charge on the ACS Publications website at DOI: 10.1021/acs.nano-lett.8b00273.

Nanosphere assembly; graphene transfer; Raman spectroscopy and strain analysis; computational methods; additional simulation results; comparison with prior work on graphene/nanoparticle systems (PDF)

## ■ AUTHOR INFORMATION

### Corresponding Author
*E-mail: nadya@illinois.edu.

### ORCID
Yingjie Zhang: 0000-0002-2704-8894
Mohammad Heiranian: 0000-0002-2227-4948
Stefano Zapperi: 0000-0001-5692-5465
Joseph W. Lyding: 0000-0001-7285-4310

### Notes
The authors declare no competing financial interest.

## ■ ACKNOWLEDGMENTS

Y.Z. was supported by a Beckman Institute Postdoctoral Fellowship at the University of Illinois at Urbana−Champaign, with funding provided by the Arnold and Mabel Beckman Foundation. N.M., N.R.A., and P.Y.H. acknowledge support from the NSF-MRSEC under Award Number DMR-1720633. Y.Z. acknowledges research support from the National Science Foundation under Grant No. ENG-1434147. J.W.L. acknowledges support from the Office of Naval Research under Grant No. N00014-16-1-3151. B.J. and P.Y.H. acknowledge support from the Air Force Office of Scientific Research under Award No. FA9550-7-1-0213. The experimental work was carried out in part in the Frederick Seitz Materials Research Laboratory Central Facilities and in the Beckman Institute at the University of Illinois. M.H. and N.R.A. acknowledge the use of the parallel computing resource Blue Waters provided by the University of Illinois and National Center for Supercomputing Applications. Z.B. and S.Z. are supported by the ERC Advanced Grant No. 291002 SIZEFFECTS.

## ■ ABBREVIATIONS

vdW, van der Waals; L-J, Lennard-Jones; Gr, graphene; NS, nanosphere; Gr-20, graphene on 20 nm nanospheres; MD, molecular dynamics

# ■ REFERENCES